\documentclass[11pt]{article}
\usepackage{amssymb, amsmath}
\usepackage{graphicx}
\newcommand{\bm}[1]{\mbox{\boldmath $#1$}}
\begin{document}
\makeatletter
\@addtoreset{equation}{section}
\def\theequation{\thesection.\arabic{equation}}
\def\@maketitle{\newpage
 \null
 {\normalsize \tt \begin{flushright} 
  \begin{tabular}[t]{l} \@date  
  \end{tabular}
 \end{flushright}}
 \begin{center} 
 \vskip 2em
 {\LARGE \@title \par} \vskip 1.5em {\large \lineskip .5em \begin{tabular}[t]{c}\@author 
 \end{tabular}\par} 
 \end{center}
 \par
 \vskip 1.5em} 
\makeatother
\topmargin=-1cm
\oddsidemargin=1.5cm
\evensidemargin=-.0cm
\textwidth=15.5cm
\textheight=22cm
\setlength{\baselineskip}{16pt}
\title{On Unimodular Gauge of Quantum Gravity}
\author{
Ryuichi~{\sc Nakayama}\thanks{nakayama@particle.sci.hokudai.ac.jp} 
       \\[1cm]
{\small
    Division of Physics, Graduate School of Science,} \\
{\small
           Hokkaido University, Sapporo 060-0810, Japan}
}
\date{
July, 2021
}
%
%
\maketitle

\begin{abstract} 
BRST-invariant action of general relativity in the unimodular gauge proposed by Baulieu is studied without using perturbative expansions. 
The  expression for the path integral in the unimodular gauge is reduced to a form in which a functional measure is defined by a norm invariant under Transverse Diffeomorphism. It is shown  that general relativity in the unimodular gauge with this action and the quantum unimodular gravity are equivalent. It is also shown that Vacuum Expectation Values (VEVs) of Diff invariant operators  in the unimodular gauge and other gauge such as the harmonic gauge take  distinct values. A path integral for harmonic gauge  is found to be gauge equivalent to a superposition of that for unimodular gauge obtained by performing constant Weyl transformation of the metric, after a non-dynamical cosmological term is introduced into the action of the  unimodular gauge.

\end{abstract}
\newpage
\setlength{\baselineskip}{18pt}
\section{Introduction}
\hspace*{5mm}
A. Einstein\cite{E} noticed existence of a special frame where volume element $\sqrt{-g}$ is fixed to unity. 
The gravity theory with a constraint $\sqrt{-g}=\omega$ is called Unimodular Gravity (UG) theory. Here $\omega$ is an arbitrary fixed volume element. In the classical version of this theory the cosmological constant is introduced as a constant of integration, which can be chosen at will for several purposes.\cite{Finkelstein} It is expected that quantum UG theory might solve the cosmological constant problem.\cite{HT}\cite{Smolin} 
At the classical level UG is not distinguishable from a gauge fixed version of General Relativity (GR), where the condition $\sqrt{-g}=\omega$ is chosen as one  of the gauge conditions. This gauge  is called unimodular gauge. At the quantum level equivalence of UG and GR is subtle\cite{Alv}\cite{Alvarez}\cite{Tinge}\cite{Padilla}.  In \cite{det1}\cite{det2} it is concluded that the effective actions at one-loop order of the  two theories are equivalent. GR with Einstein-Hilbert action  is not renormalizable. It is known that 4d GR  with extra terms in the action which are quadratic order in the Ricci tensor and Ricci scalar is renormalizable, but it is non-unitary\cite{Stelle}. There is a proposal to extend this action to make GR unitary and at the same time  super-renormalizable or even finite by promoting the coefficients of powers of Ricci tensor and Ricci scalar to entire functionals of the d'Alembertian.\cite{Tom}\cite{Modesto}\cite{Modesto2}\cite{PC}
There are also attempts to find UV attractive non-Gaussian fixed points in the renormalization group flow of UG and GR\cite{Reuter}\cite{Eichhorn}\cite{BP}\cite{det3}, and solve unitarity problem\cite{Basile}\cite{Basile2}.

In \cite{Bau1}\cite{Bau2} BRST invariant action of GR in the unimodular gauge was constructed by L. Baulieu. In GR Diffeomorphism transformation generates Weyl transformation of the metric tensor. 
To construct a BRST invariant action with unimodular gauge fixing in addition to harmonic gauge fixing it is necessary to introduce two BRST quartets and arrange BRST-invariant terms so as to avoid over-gauge fixing and singularities of the propagators of the fields in the BRST quartets. 

One of the purposes of this paper is to rewrite the BRST invariant action found in \cite{Bau1} into the same form as that of the action for UG and transform the functional measure for the path integral into a form invariant under the Transverse Diffeomorphism group. By studying the functional measure of the path integral it is found that if the diffeomorphism invariant measures are rewritten in terms of a measure invariant under transverse diffeomorphism. It will be shown that the path integral of GR in the unimodular gauge ($\sqrt{-g}=\omega$) coincide with that of  UG. 
This is shown in sec.2. 

Difference between unimodular gauge and other gauges of GR lies in the treatment of the cosmological term. In UG a volume element $\sqrt{-g}$ is fixed to $\omega$ in any gauge, because the conformal mode is absent in UG, and the cosmological term does not contribute to dynamics. On the other hand in GR in general gauges, except for the unimodular gauge, the volume element is not fixed and the cosmological constant will play a role of a self-coupling constant of gravitation. Physical quantities will depend on the cosmological constant. In the unimodular gauge, however, the cosmological term is decoupled from the other terms of the action. Hence in the unimodular gauge vacuum expectation values (VEVs) of physical operators, which are invariant under diffeomorphism,  will not depend on the cosmological constant. In GR, however, the gauge fixing can be changed by adding BRST-exact terms to the action. 
So a question arises as to  how VEVs of physical operators in gauges other than unimodular gauge are related to those in the unimodular gauge. The second purpose of this paper is to study this problem. 

 In sec.3 we will add an extra BRST-exact term 
\begin{equation}
\beta \int d^d x \, \omega \,  s(\eta \, L), \label{per}
\end{equation}
where $\eta$ and $L$ are fields of a BRST quartet and $s$ a graded generator of BRST transformation\cite{Bau1},  to the action for GR in the unimodular gauge.  The new action contains a constant gauge parameter $\beta$. For $\beta \rightarrow 0$ the action is expected to coincide with that for unimodular gauge. For $\beta \rightarrow \infty$ the theory coincides with GR in the harmonic gauge. It turnes out, however, in the $\beta \rightarrow 0$ limit the theory reduces to  GR in a new  gauge $\sqrt{-g}=\mu \omega$, where $\mu$ is a positive constant, over which  integration must be carried out in the path integral. 
VEVs of physical operators in this interpolating gauge depend on the cosmological constant. So VEVs of physical operators in the unimodular gauge and in other gauges will not coincide. 
It is shown that a path integral for harmonic gauge of GR is gauge equivalent to a superposition of those for unimodular gauge  obtained by performing  constant  Weyl transformation $\hat{g}_{\mu\nu} \rightarrow \mu^{2/d} \hat{g}_{\mu\nu}$, after a non-dynamical cosmological term is introduced into the action of unimodular gauge. An integration measure for $\mu$ is proposed. 
In sec. 4 summary and discussions are presented.

\section{Relation between Path Integrals for  General Relativity in the Unimodular Gauge and Unimodular Gravity}
\hspace*{5mm}
In \cite{Bau1} Baulieu constructed a BRST-invariant gauge fixed action for Einstein gravity in the unimodular gauge. This gauge is defined by
\begin{equation}
\sqrt{-g}=\omega, \qquad (g=\text{det} \, g_{\mu\nu})  \label{unigauge}
\end{equation}
and 
\begin{equation}
\partial_{\nu}\hat{g}^{\nu \mu}=0, \qquad (\mu,\nu=0,1,2, \ldots, d-1)
\end{equation}
where $\hat{g}_{\mu\nu}$ is related to $g_{\mu\nu}$ by $g_{\mu\nu}=(\sqrt{-g}/\omega)^{2/d}
 \hat{g}_{\mu\nu}$. $\omega(x)$ is a fixed volume element of the spacetime.  
Gauge fixed action\cite{Bau1} is given by\footnote{Here  the volume element $\omega$ and a cosmological constant $\Lambda$ are introduced to the action obtained in \cite{Bau1}. This term will be necessary for studying connections of the unimodular gauge with the harmonic gauge in later sections. In this unimodular gauge the cosmological term is non-dynamical. }  
\begin{equation}
S_{\text{BRST}}=\frac{1}{16\pi G}\int d^dx \Big( \sqrt{-g}(R(g)-2\Lambda)+\omega s \Big[\overline{\xi}^{\mu}(\hat{g}_{\mu\nu}\partial_{\rho}\hat{g}^{\rho\nu}+\gamma \partial_{\mu}L+\frac{1}{2}\alpha \hat{g}_{\mu\nu}b^{\nu})+\overline{\eta}(\sqrt{-g}/\omega-1)\Big]\Big) \label{S1}
\end{equation}
Here $\alpha$ and $\gamma (\neq 0)$ are constants. $R(g)$ is a Ricci scalar. Matter action is not written explicitly for simplicity. 
There are two sets of BRST systems. One is composed of $g_{\mu\nu}$, $\xi^{\mu}$, $\overline{\xi}^{\mu}$ and $b^{\mu}$. This multiplet is associated with diffeomorphism. The graded differential operator $s$ generates BRST transformation on this multiplet as follows.
\begin{eqnarray}
s \, g_{\mu\nu} &=& g_{\mu\rho} \partial_{\nu} \xi^{\rho}+g_{\nu\rho} \partial_{\mu} \xi^{\rho}+\xi^{\rho}\partial_{\rho} g_{\mu\nu}, \nonumber \\
s \, \xi^{\mu} &=& \xi^{\nu}\partial_{\nu}\xi^{\mu}, \nonumber \\
s \, \overline{\xi}^{\mu} &=& b^{\mu}, \nonumber \\
s \, b^{\mu} &=& 0.  \label{BRS1}
\end{eqnarray}
Here $\xi^{\mu}$ and $\overline{\xi}^{\mu}$ are ghost and anti-ghost fields. $b^{\mu}$ is a Nakanishi-Lautrup B-field. 
The other multiplet is given by $L$, $\eta$, $\overline{\eta}$ and $b$. BRST transformation is given by
\begin{eqnarray}
s \, L &=& \eta, \nonumber \\
s \, \eta &=& 0, \nonumber \\
s \, \overline{\eta} &=& b, \nonumber \\
s \, b &=& 0. \label{BRS2}
\end{eqnarray}
Here $\eta$ and $\overline{\eta}$ are ghost and anti-ghost. 
The above two BRST transformations are nilpotent. 

After BRST transformations in the integrand are worked out in (\ref{S1}) we obtain
\begin{eqnarray}
S_{\text{BRST}} &=&\frac{1}{16\pi G}\int d^dx \Big[ \sqrt{-g}\, (R(g)-2\Lambda)+\omega\,  \Big\{b^{\mu}(\hat{g}_{\mu\nu}\partial_{\rho}\hat{g}^{\rho\nu}+\gamma \, \partial_{\mu}L+\frac{1}{2}\alpha \, \hat{g}_{\mu\nu}b^{\nu})\nonumber \\
&&-\frac{1}{2}\alpha \, \overline{\xi}^{\mu}s \, (\hat{g}_{\mu\nu})b^{\nu}+b\, (\sqrt{-g}/\omega-1) \nonumber \\
&& -\overline{\xi}^{\mu} s \,(\hat{g}_{\mu\nu}\partial_{\rho}  \hat{g}^{\rho\nu})-\gamma \, \overline{\xi}^{\mu}\partial_{\mu}\eta+\overline{\eta}\nabla_{\nu}\xi^{\nu}\Big\}\Big]
\end{eqnarray}
Here we note that 
\begin{eqnarray}
s\, \hat{g}_{\mu\nu} &=& \hat{g}_{\mu\rho}\partial_{\nu}\xi^{\rho}+\hat{g}_{\nu\rho}\partial_{\mu}\xi^{\rho}+\xi^{\rho}\partial_{\rho}\hat{g}_{\mu\nu}-\frac{2}{d}\hat{g}_{\mu\nu}\hat{\nabla}_{\rho} \xi^{\rho}, \\
\hat{\nabla}_{\rho}\xi^{\rho} &=& \frac{1}{\omega}\partial_{\rho}(\omega \xi^{\rho}), \\
s \, \sqrt{-g} &=& \sqrt{-g} \, \nabla_{\nu}\xi^{\nu}
\end{eqnarray}
$\nabla_{\mu}$ and $\hat{\nabla}_{\mu}$ are covariant derivatives associated with $g_{\mu\nu}$ and $\hat{g}_{\mu\nu}$, respectively. 
The equation of motion for $b$ imposes $\sqrt{-g}=\omega$, {\it i.e.}, ($g_{\mu\nu}=\hat{g}_{\mu\nu}$). 
Elimination of $b$ replaces the first term in the integrand by $\displaystyle \sqrt{-\hat{g}} (R(\hat{g})-2\Lambda)$ and 
an action integral in the unimodular gauge is obtained. Also $\overline{\eta}\nabla_{\nu}\xi^{\nu}$ is replaced by $\overline{\eta}\hat{\nabla}_{\nu}\xi^{\nu}= \frac{1}{\omega}\overline{\eta}\partial_{\nu}(\omega \xi^{\nu})$. 

The field $L$ plays the role of balancing the degrees of freedom.\cite{Bau1}  Because the unimodular condition $\sqrt{-g}=\omega$ fixes a single component of the metric tensor, d-1 more conditions can be imposed. Now, 
$\hat{g}_{\mu\nu}\partial_{\rho}\hat{g}^{\rho\nu}$ has d components. So addition of $\gamma \partial_{\mu}L$ and integration of $L$ over $(-\infty, \infty)$ avoids over-gauge fixing. 

This field $L$ also plays another important role: Functional integration over $L$ forces $\hat{\nabla}_{\mu}b^{\mu}=0$. Further integration over $\eta$ and $\overline{\eta}$ also forces conditions $\hat{\nabla}_{\mu} \xi^{\mu}=\hat{\nabla}_{\mu}\overline{\xi}^{\mu}=0$. As is clear from (\ref{BRS1}) this means that the BRST multiplet $(\hat{g}_{\mu\nu}, b^{\mu}, \xi^{\mu}, \overline{\xi}^{\mu})$ is reduced to that for transverse diffeomorphism (TDiff), or in other words, volume-preserving diffeomorphism. After the reduction the BRST invariant action  $S_{\text{reduced}}$ is given by
\begin{eqnarray}
S_{\text{reduced}}[\hat{g}_{\mu\nu},\xi^{\mu},\bar{\xi}^{\mu},b^{\mu};\omega] &=& \frac{1}{16\pi G}\int d^dx \,  \omega \, \Big[ (R(\hat{g})-2\Lambda)  + b^{\mu}(\hat{g}_{\mu\nu}\partial_{\rho}\hat{g}^{\rho\nu})+\frac{1}{2}\alpha \hat{g}_{\mu\nu}b^{\mu}b^{\nu} \nonumber \\
&& \qquad -\frac{1}{2}\alpha\overline{\xi}^{\mu}s \, (\hat{g}_{\mu\nu})b^{\nu}-\overline{\xi}^{\mu}s \, (\hat{g}_{\mu\nu}\partial_{\rho}\hat{g}^{\rho\nu})\Big]  \label{S55}
\end{eqnarray}
This action coincides with the BRST invariant action of quantum UG (QUG) with a condition $\text{det}\hat{g}_{\mu\nu}=\hat{g}=-\omega^2$ except for the cosmological term.\footnote{In UG there is no cosmological term.  } 
\begin{eqnarray}
S_{\text{QUG}}[\hat{g}_{\mu\nu},\xi^{\mu},\bar{\xi}^{\mu},b^{\mu}:\omega]  &=& \frac{1}{16\pi G}\int d^dx \,  \omega \, \Big[ R(\hat{g})  + b^{\mu}(\hat{g}_{\mu\nu}\partial_{\rho}\hat{g}^{\rho\nu})+\frac{1}{2}\alpha \hat{g}_{\mu\nu}b^{\mu}b^{\nu} \nonumber \\
&& \qquad -\frac{1}{2}\alpha\overline{\xi}^{\mu}s \, (\hat{g}_{\mu\nu})b^{\nu}-\overline{\xi}^{\mu}s \, (\hat{g}_{\mu\nu}\partial_{\rho}\hat{g}^{\rho\nu})\Big]  \label{S66}
\end{eqnarray}

Path integral for unimodular gauge of quantum general relativity (QGR)  is expressed as
\begin{equation}
Z_{\text{QGR, unimodular gauge}}=\int ({\cal D}g_{\mu\nu}{\cal D}b^{\mu}{\cal D}\xi^{\mu}{\cal D}\overline{\xi}^{\mu})_D \ e^{iS_{\text{reduced}}} \label{PathQGR}
\end{equation}
Integrations over $L$, $b$, $\eta$ and $\overline{\eta}$ do not create new determinants.  
In this formula the functional measures are defined via norms which are invariant under general diffeomorphism (Diff), because unimodular gauge is one of gauge fixing of GR. The subscript $D$ stands for Diff. 
Because the fields $b^{\mu}$, $\xi^{\mu}$ and $\overline{\xi}^{\mu}$ in (\ref{S55}) satisfy transversality conditions ({\em i.e.} $\hat{\nabla}_{\mu}b^{\mu}=0$, etc) and $\hat{g}_{\mu\nu}$ is substituted into $g_{\mu\nu}$, these measures must be rewritten in terms of  the ones defined by TDiff-invariant norm. 

Path integral in QUG is, however, given by 
\begin{equation}
Z_{\text{QUG}}=\int ({\cal D}\hat{g}_{\mu\nu}{\cal D}b^{\mu}{\cal D}\xi^{\mu}{\cal D}\overline{\xi}^{\mu})_T \ e^{iS_{\text{QUG}}} \label{PathQUG}
\end{equation}
Here the subscript $T$ on the measure means that this measure is defined by TDiff -invariant norm. 

We first consider $({\cal D} b^{\mu})_D$. The measure for this Nakanishi-Lautrup field was first defined for genaral diffeomorphism group. 
For general diffeomorphism an infinitesimal vector $\delta b^{\mu}$ can be decomposed into a transverse part $\delta b_T^{\mu}$ and a longitudinal part.
\begin{equation}
\delta b^{\mu}(x)= \delta b_T^{\mu}(x)+ \hat{\nabla}^{\mu}\phi_b(x),   \label{bexp}
\end{equation}
The transverse part satisfies $\hat{\nabla}_{\mu} \delta b^{\mu}_T=0$. The scalar field $\phi_b$ does not appear in the action in the unimodular gauge. The integration measure $({\cal D}_{\hat{g}}b^{\mu})_{D}$ is defined using the $L^2$ norm with respect to the metric $\hat{g}$ as \cite{DK}
\begin{equation}
\int ({\cal D}_{\hat{g}} \delta b^{\mu})_D \exp (-\parallel \delta b^{\mu} \parallel^2_{\hat{g}   })=1, 
\end{equation}
where
\begin{equation}
\parallel \delta b^{\mu} \parallel^2_{\hat{g} } \, =\int d^dx \, \omega \hat{g}_{\mu\nu} \, \delta b^{\mu}\delta b^{\nu}
\end{equation}
Substituting (\ref{bexp}) into the above yields 
\begin{equation}
\parallel \delta b^{\mu} \parallel^2_{\hat{g} } \, =\int d^dx \, \omega \hat{g}_{\mu\nu} \, \delta b_T^{\mu}\delta b_T^{\nu}+ \int d^d x \omega \phi_b (-\hat{\nabla}^2\phi_b)  \label{deltab}
\end{equation} 
So up to an infinite multiplicative constant, which is independent of $\hat{g}_{\mu\nu}$,  we obtain
\begin{equation}
({\cal D}b^{\mu})_D=({\cal D}b^{\mu})_T\,  \big(\text{Det} (-\hat{\nabla}^2)\big)^{\frac{1}{2}} \times {\cal D}\phi_b. 
\end{equation}
This contains a determinant of a Laplacian. In (\ref{bexp}) there is a first order derivative of $\phi_b$. By a transformation of the variable, $\phi_b \rightarrow \tilde{\phi}_b= (-\hat{\nabla}^2)^{1/2} \phi_b$, we have  $\hat{\nabla}^{\mu} \phi_b=\hat{\nabla}^{\mu}(-\hat{\nabla}^2))^{-1/2}\tilde{\phi}_b$ and  the measure will be rewritten as 
\begin{equation}
({\cal D}b^{\mu})_D=({\cal D}b^{\mu})_T\, {\cal D}\tilde{\phi}_b. \label{medb}
\end{equation}
Because the action (\ref{S55}) does not depend on the field $\tilde{\phi}_b$,  integral $\int {\cal D}\tilde{\phi}_b$ can be dropped, being an infinite constant.
$({\cal D}b^{\mu})_T$ stands for a measure for transverse vectors and defined by a norm invariant under  TDiff, while $({\cal D}b^{\mu})_D$ is a measure in QGR which is defined with respect to a norm invariant under general diffeomorphism (Diff). Similar notation will be used for other measures. 
Here the subscript $\hat{g}$ is omitted in all the factors. The measures $({\cal D}\xi^{\mu})_D$ and $({\cal D}\overline{\xi}^{\mu})_D$ can be treated in a similar fashion, except for the fact that $\xi^{\mu}$ and $\overline{\xi}^{\mu}$ are Grassmann-odd fields. So the results are given by
\begin{eqnarray}
({\cal D} \xi^{\mu})_D&=& ({\cal D}\xi^{\mu})_T\,  \big(\text{Det} (-\hat{\nabla}^2)\big)^{-\frac{1}{2}}{\cal D}\phi_{\xi}=({\cal D}\xi^{\mu})_T\,  {\cal D}\tilde{\phi}_{\xi}, \label{medxi} \\
({\cal D} \hat{\xi}^{\mu})_D&=& ({\cal D}\overline{\xi}^{\mu})_T \,  \big(\text{Det} (-\hat{\nabla}^2)\big)^{-\frac{1}{2}}{\cal D}\phi_{\overline{\xi}}=({\cal D}\overline{\xi}^{\mu})_T{\cal D}\tilde{\phi}_{\overline{\xi}}. \label{medxib}
\end{eqnarray}
Here, ghost $\xi^{\mu}$ is decomposed into transverse and longitudinal parts as $\xi^{\mu}=\xi^{\mu}_T+\hat{\nabla}^{\mu}\phi_{\xi}$. $\phi_{\xi}$ is redefined such that  ${\cal D}\tilde{\phi}_{\xi}=(\text{Det}(-\hat{\nabla}^2))^{-1/2}{\cal D}\phi_{\xi}$. 
Similarly for the measure of $\overline{\xi}^{\mu}$. In the unimodular gauge $\xi^{\mu}$ and $\overline{\xi}^{\mu}$ are transverse, and the action does not 
depend on $\tilde{\phi}_{\xi}$ and $\tilde{\phi}_{\overline{\xi}}$. Hence  integral over these fields are constants. Because  $\tilde{\phi}_{\xi}$ and $\tilde{\phi}_{\overline{\xi}}$ are Grassmann odd variables, these constants actually vanish. They will be canceled by the infinite integral over $\tilde{\phi}_b$ and a similar integral over $\tilde{\phi}_g$ which also appears below in (\ref{medg}). 

Analysis for measure ${\cal D}g_{\mu\nu}$ is similarly carried out. 
This is carried out as follows. 
Fluctuation $\delta \hat{g}_{\mu\nu}$ is traceless: $h \equiv \hat{g}^{\mu\nu} \delta \hat{g}_{\mu\nu}=0$ due to the unimodular condition. $h$ does not appear in the action. 
In the measure  fluctuation of $g_{\mu\nu}$ around $\hat{g}_{\mu\nu}$, $\delta g_{\mu\nu}=g_{\mu\nu}-\hat{g}_{\mu\nu}$, is decomposed into York variables.\cite{det1} 
\begin{equation}
\delta g_{\mu\nu}= h^{TT}_{\mu\nu}+(\hat{\nabla}_{\mu}\zeta_{T\nu}+\hat{\nabla}_{\nu}\zeta_{T\mu})+(\hat{\nabla}_{\mu}\hat{\nabla}_{\nu}-\frac{1}{d}\hat{g}_{\mu\nu}\hat{\nabla}^2)\sigma+\frac{1}{d}\hat{g}_{\mu\nu} \, h,  \label{deltagmn}
\end{equation}
where $\hat{\nabla}^{\mu}h^{TT}_{\mu\nu}=0$, $\hat{g}^{\mu\nu}h^{TT}_{\mu\nu}=0$, $\hat{\nabla}^{\mu}\zeta_{T\mu}=0$, $h=\hat{g}^{\mu\nu}\delta g_{\mu\nu}$. 
In this case the functional measure is defined by a norm
\begin{equation}
\parallel \delta g_{\mu\nu} \parallel^2_{\hat{g} } \, =\int d^dx \, \omega \, \hat{g}^{\mu\nu} \hat{g}^{\lambda\rho}\, \big[\delta g_{\mu\lambda} \delta g_{\nu\rho}+\lambda \delta g_{\mu\nu} \delta g_{\lambda\rho}\big] \label{gnorm}
\end{equation}
Here $\lambda$ is an arbitrary positive constant. 
In the case of Diff, where  the infinitesimal diffeomorphism parameters are given by
\begin{equation}
 \epsilon^{\mu}=\epsilon_T^{\mu}+\hat{\nabla}^{\mu}\phi_g, \qquad \hat{\nabla}_{\mu}\epsilon_T^{\mu}=0, 
\end{equation}
it can be shown \cite{det1} that the transformation properties of the York variables are 
\begin{equation}
 \zeta_T^{\mu}=\epsilon_T^{\mu}, \qquad h=2\hat{\nabla}^2 \phi_g, \qquad \sigma=2\phi_g, \qquad \delta h^{TT}_{\mu\nu}=0.   \label{param}
\end{equation} 
Then the norm of $\delta g_{\mu\nu}$ is given by
\begin{eqnarray}
\parallel \delta g_{\mu\nu} \parallel^2_{\hat{g} } \, =4(1+\lambda) \int d^d x \omega \, [(-\hat{\nabla}^2) \phi_g]^2 
+2\int d^dx \omega (\hat{\nabla}_{\mu}\epsilon_{T\lambda})[ \hat{\nabla}^{\mu}\epsilon_T^{\lambda}+  \hat{\nabla}^{\lambda}\epsilon_T^{\mu}].    \label{del g}
\end{eqnarray} 
Because difference between $({\cal D}g_{\mu\nu})_D$ and $({\cal D}g_{\mu\nu})_T$ is a part which depends on $\phi_g$, the measure can be rewritten as 
\begin{equation}
({\cal D}g_{\mu\nu})_D=({\cal D}\hat{g}_{\mu\nu})_T \, \text{Det} (-\hat{\nabla}^2) \, {\cal D}\phi_g=
({\cal D}\hat{g}_{\mu\nu})_T {\cal D}\tilde{\phi}_g
 \label{medg}
\end{equation}
Here $\phi_g$ is redefined as $\tilde{\phi}_g=(-\hat{\nabla}^2)\phi_g$, because there is a term proportional to  $h = 2\hat{\nabla}^2 \phi_g$ in (\ref{deltagmn}). 
Because $h=0$ holds in the action and the action does not depend on $\tilde{\phi}_g$ due to relation $h=2\hat{\nabla}^2 \phi_g$ from (\ref{param}),  
path integral over $\tilde{\phi}_g$ yields  an infinite constant and can be discarded. 
Combining the above results, (\ref{medb}), (\ref{medxi}), (\ref{medxib}) and  (\ref{medg}), we finally obtain
\begin{equation}
({\cal D}g_{\mu\nu}
{\cal D}b^{\mu}{\cal D}\xi^{\mu}{\cal D}\overline{\xi}^{\mu} )_D
=({\cal D}\hat{g}_{\mu\nu}{\cal D}b^{\mu}{\cal D}\xi^{\mu}{\cal D}\overline{\xi}^{\mu})_T. 
\end{equation}
To summarize, path integral formula (\ref{PathQGR}) for QGR in the unimodular gauge is rewritten as
\begin{equation}
Z_{\text{QGR, unimodular gauge}}=\int ({\cal D}\hat{g}_{\mu\nu}{\cal D}b^{\mu}{\cal D}\xi^{\mu}{\cal D}\overline{\xi}^{\mu})_T \ e^{iS_{\text{reduced}}}  \label{ZBRSTp}
\end{equation}
This path integral coincides with (\ref{PathQUG}) for the path integral in QUG. 

In \cite{det1} and \cite{det2} it was concluded that the effective actions at one-loop order in UG and  GR  in a gauge different from the unomodular gauge of \cite{Bau1}, coincide. 

Relation between the path integrals in the unimodular gauge and the harmonic gauge of GR will be studied in the next section. 

\section{Action Interpolating between Unimodular Gauge and Harmonic Gauge}
\hspace*{5mm}
 The cosmological term in the unimodular gauge of GR does not couple to other terms in the action.  So a question arises whether the action (\ref{S1}) is related to that of other gauges of GR such as harmonic gauge, if additional appropriate BRST-exact terms are added to the action. 

In this section we will consider an action integral which is expected to interpolate between  GR in the unimodular gauge and GR in the harmonic gauge, and study the `unimodular gauge limit' of the action. 

\subsection{Action in the Interpolating gauge}
\hspace{5mm}
The action integral (\ref{S1}) is replaced by 
\begin{eqnarray}
S_{\text{Interpolating}}&=&\frac{1}{16\pi G}\int d^dx  \, \omega \Big(\omega^{-1}\sqrt{-g}(R(g)-2\Lambda )+  s \Big[\overline{\xi}^{\mu}(g_{\mu\nu}\partial_{\rho}g^{\rho\nu}+\gamma \, \partial_{\mu}L+\frac{1}{2}\alpha \, g_{\mu\nu} \, b^{\nu}) \nonumber \\
&&+ \overline{\eta} \, (\sqrt{-g}/\omega-1)\Big]\Big)+\frac{1}{16\pi G}\int d^dx  \, \omega \,
s\Big(\beta \, \overline{\eta} \, L\Big)   \label{S2}
\end{eqnarray}
Difference between $S_{\text{Interpolating}}$ and $S_{\text{BRST}}$ (\ref{S1})  is that  the last term $s (\beta \overline{\eta}L)$ in the second line is added and that $\hat{g}_{\mu\nu}$'s in the gauge fixing term is replaced by $g_{\mu\nu}$'s.  When  $\beta=0$ exactly, although $g_{\mu\nu}$ and $g^{\rho \nu}$ are used instead of $\hat{g}_{\mu\nu}$ and $\hat{g}^{\rho\nu}$ in the gauge fixing terms of  (\ref{S2}), these terms are equivalent to the corresponding terms in (\ref{S1}) due to the constraint $\sqrt{-g}/\omega-1=0$, which is enforced after integration over $b=s\overline{\eta}$. ($\sqrt{-g}=\omega$ implies $g_{\mu\nu}=\hat{g}_{\mu\nu}$.) So as $\beta \rightarrow 0$, action (\ref{S2}) is expected to  reduce to (\ref{S1}). This non-unimodular gauge parametrized by $\beta$ will be henceforth called an `interpolating gauge'. 

When $\beta \neq 0$ and $\alpha=0$, the gauge conditions are 
\begin{eqnarray}
&& g_{\mu\nu} \, \partial_{\rho}g^{\rho\nu}+\gamma \, \partial_{\mu}L=0, \\
&& \sqrt{-g}/\omega-1=-\beta L.
\end{eqnarray}
There are $d+1$ conditions. 
So $L$ is determined as $L=(1/\beta)(1-\sqrt{-g}/\omega)$ and the gauge condition on the metric is given by
\begin{equation}
g_{\mu\nu}\partial_{\rho}g^{\rho\nu}=\frac{\gamma}{\beta} \, \partial_{\mu} \big(\frac{\sqrt{-g}}{\omega}\big).
\end{equation}

For general $\beta$, after carrying out BRST transformation inside the integrand of (\ref{S2}) the action takes the following form. 
\begin{eqnarray}
S_{\text{Interpolating}}&=&\frac{1}{16\pi G}\int d^dx  \, \omega \, \Big(\omega^{-1}\sqrt{-g}(R(g)-2\Lambda) + b^{\mu} (g_{\mu\nu}\partial_{\rho} g^{\rho \nu}+\gamma\partial_{\mu} L)+\frac{1}{2}\alpha g_{\mu\nu}b^{\mu}b^{\nu}
\nonumber \\
&&- \overline{\xi}^{\mu}s \, (g_{\mu\nu}\partial_{\rho}g^{\rho\nu})-\gamma  \overline{\xi}^{\mu}\partial_{\mu}\eta+ b(\frac{\sqrt{-g}}{\omega}+\beta L-1)- \overline{\eta}\nabla_{\nu}\xi^{\nu}- \beta \overline{\eta}\eta \nonumber \\
&& -\frac{1}{2} \alpha  \overline{\xi}^{\mu}b^{\nu}s \, g_{\mu\nu}
\Big)
\label{S3}
\end{eqnarray}

 By integrating  over $b$, a delta function $\delta (\sqrt{-g}/\omega+\beta L-1)$ is obtained. 
Then after integration over $L$, $\eta$ and $\overline{\eta}$ the action integral is given by
\begin{eqnarray}
S_{\text{Interpolating}}&=&\frac{1}{16\pi G}\int d^dx\,  \omega \,  \Big(\omega^{-1}\sqrt{-g}(R(g)-2\Lambda) + b^{\mu} \{g_{\mu\nu}\partial_{\rho} g^{\rho \nu}-\frac{\gamma}{\beta}\partial_{\mu} (\sqrt{-g}/\omega)\}\nonumber \\
&&+\frac{1}{2}\alpha  g_{\mu\nu}b^{\mu}b^{\nu}
- \overline{\xi}^{\mu}s \, (g_{\mu\nu}\partial_{\rho}g^{\rho\nu})
-\frac{\gamma}{\beta}  \nabla_{\mu} \overline{\xi}^{\mu} \hat{\nabla}_{\nu}\xi^{\nu}       
-\frac{1}{2} \alpha \overline{\xi}^{\mu}b^{\nu} (s \, g_{\mu\nu}) \Big)
\label{S4}
\end{eqnarray}
In the limit $\beta \rightarrow \infty$ the action will reduce to that in the `harmonic gauge', where $\sqrt{-g}$ in the terms except for $\sqrt{-g}(R(g)-2\Lambda)$ are replaced by $\omega$. Furthermore, by adding 
\begin{equation}
S_{\text{additional BRS exact term}}=
\frac{1}{16\pi G}\int d^dx\, s\Big[(\sqrt{-g}-\omega) \Big\{\frac{1}{2}\alpha \overline{\xi}^{\mu}g_{\mu\nu}b^{\nu} +\overline{\xi}^{\mu}g_{\mu\nu}\partial_{\rho}g^{\rho \nu}  \Big\}      \Big]
\end{equation}
to the action, the gauge choice will become the genuine harmonic gauge. For simplicity this additional BRST-exact term will be omitted in the following investigation. 

As a limit  $\beta \rightarrow 0$ is taken in (\ref{S4}), we obtain the relations 
\begin{equation}
\sqrt{-g}/\omega=\mu (=\text{const}),  \qquad \hat{\nabla}_{\mu}\xi^{\mu}=0, \qquad \nabla_{\mu}\overline{\xi}^{\mu}=0
\end{equation}
due to the terms with the coefficient $\gamma/\beta$. 
 So in this limit the BRST quartet (\ref{BRS1}) will reduce to that for TDiff symmetry and the integration measure can be expressed in terms of that for QGR based on TDiff-invariant norm as in sec 2.\footnote{Due to the first condition $\sqrt{-g}=\mu \omega=\sqrt{-\hat{g}}$. So $\nabla_{\mu}\overline{\xi}^{\mu}=\hat{\nabla}_{\mu}\overline{\xi}^{\mu} $.} 
On the other hand $\mu$ is an arbitrary positive constant and this scale factor $\mu$ is left undetermined. 
The behavior of the path integral at $\beta \rightarrow 0$ changed from that in the case $\beta=0$, because the field $L$ is an integration variable in the path integral and $|\beta L|$ can take large values, even if limit $\beta \rightarrow 0$ is taken.

For any choice of $\mu$ the action coincides with that for unimodular gauge  with a new fixed volume element $\mu\omega$. Because $\sqrt{-g}=\mu\omega$ means $g_{\mu\nu}=\mu^{2/d}\hat{g}_{\mu\nu}$, some terms in the BRST invariant action depend on $\mu$.  It is necessary to carry out integration over $\mu$ in the path integral, because $\mu$ is a constant mode of the  variable $L$. It will not be natural to select a special value $\mu=1$. In the next subsection it will be shown that $\mu=1$ cannot be chosen, because there must be a discontinuity of VEVs at $\beta=0$. 
Furthermore there will be no natural prescription to choose other single value of $\mu$.

\subsection{Discontinuity of VEVs at $\beta=0$}
\hspace{5mm}
If the vacuum is invariant under BRST transformations, VEVs of physical operators $ O=  \int d^d x \sqrt{-g} \, {\cal O}(x) $, {\em i.e.}, the operators invariant under diffeomorphisms, will not usually depend on the gauge choices.  

VEV of $O$ in the unimodular gauge, $\langle 0| O |0\rangle_{\text{unimodular}}\equiv \int d^dx \omega {\cal O}(x)$ does not depend on the cosmological constant, because in this gauge the cosmological term is not a dynamical field and can be put outside the path integral. On the other hand VEV of $O$ in the interpolating gauge defined in subsec.3.1, $\langle 0| O |0\rangle_{\text{Interpolating}}(\Lambda)$, depends on the cosmological constant $\Lambda$. Furthermore, $\langle 0| O |0\rangle_{\text{Interpolating}}$ does not depend on the gauge parameter $\beta$, because the part of the action, which is proportional to $\beta$, is BRST-exact. Then we obtain
for the VEV in the harmonic gauge
\begin{equation}
\langle 0| O |0\rangle_{\text{harmonic}}= \langle 0| O |0\rangle_{\text{Interpolating}}    (\Lambda)
\end{equation}
This cannot be equal to $\langle 0| O |0\rangle_{\text{unimodular}}$, which is independent of $\Lambda$. 
\begin{equation}
\langle 0| O |0\rangle_{\text{harmonic}} =\langle 0| O |0\rangle_{\text{Interpolating}}(\Lambda)    \neq \langle 0| O |0\rangle_{\text{unimodular}}
\end{equation}
This shows that there must be a discontinuity in VEVs at $\beta=0$. If only the value $\mu=1$ were selected in the $\beta \rightarrow 0$ limit, then $\langle 0| O |0\rangle_{\text{harmonic}} =\langle 0| O |0\rangle_{\text{unimodular}}$ would be obtained. If a value of $\mu$ is undetermined, VEVs of diffeomorphism invariant operators would depend on $\mu$ and  be undetermined in the interpolating gauge, although they must be independent of $\beta (\neq 0)$. 
This shows that the integration over $\mu$ must be carried out. This discontinuity at $\beta=0$ is natural, because the volume of spacetime in the unimodular gauge is fixed, although the one in the interpolating gauge can take arbitrary values.

\subsection{Action and Path Integral in the $\beta \rightarrow 0$ Limit}
\hspace*{5mm}
To carry out integration over $\mu$ it is necessary to determine the integration measure for $\mu$. 
Parameter $\mu$ is a constant mode of  $L$. The measure for $\mu$ is expected to be defined by a norm for $L$.
\begin{equation}
||\delta L||_{g}^2 = \int d^dx \sqrt{-g} (\delta L)^2
\end{equation}
Because $L= \beta^{-1}(1-\mu)$, substitution of  $\delta L=-\beta^{-1} \delta \mu$ yields 
\begin{equation}
||\delta \mu||^2_{g}= \beta^{-2} \, \Big[\int d^d x \sqrt{-g} \Big]\delta \mu^2
\end{equation}
By using $\sqrt{-g}=\mu \sqrt{-\hat{g}}$ we obtain $||\delta \mu||_{g}^2=\mu ||\delta \mu||_{\hat{g}}^2$ and the measure for $\mu$ will be given by\footnote{Because $\mu$ is related to the rescaling of the metric, it is interesting to check a measure related to the norm of $\hat{g}_{\mu\nu}$. 
If we assume that  the norm for the metric tensor is chosen to be of the form (\ref{gnorm}), 
$||\delta g_{\mu\nu}||_{g}^2= \int d^d x \sqrt{-g} g^{\mu\nu}g^{\lambda\rho}(\delta g_{\mu\lambda} \delta g_{\nu\rho})$ with $ g_{\mu\nu}=\mu^{2/d}\hat{g}_{\mu\nu}$,
then we find $||\delta g_{\mu\nu}||_{g}^2 \propto \frac{1}{\mu}(\delta \mu)^2$
and the integration measure for $\mu$ would be $(1/\sqrt{\mu}) \, d\mu$.
This does not coincide with the above result determined by the norm for $L$.  By taking into account the connection of $\mu$ with $L$ the  measure (\ref{meadmu}) will be used in the following. }
\begin{equation}
\sqrt{\mu} \, d\mu.       \label{meadmu}
\end{equation}

Functional measures of gravity and matter fields depend on $\hat{g}_{\mu\nu}$ and under Weyl rescaling $\hat{g}_{\mu\nu} \rightarrow \mu^{2/d}\hat{g}_{\mu\nu}$ these functional measures will acquire extra powers of $\mu$. 
For example, because 
\begin{equation}
||\delta b^{\mu}||^2_{\hat{g}}=\int d^d x \omega \hat{g}_{\mu\nu}\delta b^{\mu}\delta b^{\nu}
\rightarrow \mu^{(2+d)/d} ||\delta b^{\mu}||^2_{\hat{g}}, 
\end{equation}
the transformation
\begin{equation}
{\cal D} b^{\mu} \rightarrow \underset{x} {\Pi} \Big( \mu^{\frac{d+2}{2d}}\Big) {\cal D}b^{\mu}
\end{equation}
is obtained. By combining the transformations of other measures and the determinant of the scalar Laplacian, we obtain the following result. 
\begin{equation}
({\cal D}\hat{g}_{\mu\nu}{\cal D}b^{\mu}{\cal D}\xi^{\mu}{\cal D}\overline{\xi}^{\mu})_T \rightarrow 
({\cal D}\hat{g}_{\mu\nu}{\cal D}b^{\mu}{\cal D}\xi^{\mu}{\cal D}\overline{\xi}^{\mu})_T  \Big[\underset{x} {\Pi}\,  \mu \Big]
\end{equation}
If matter fields exist, there will also be corresponding factors. As is usually done in the case of dimensional regularization, we will set  $\Pi_x \mu=1$ in this paper. 

The integral over $\mu$ will induce couplings between the cosmological constant term and other terms in the action and operators in the integrand of the path integral. Due to this integration over $\mu$, VEVs in the limit $\beta \rightarrow 0$ does not coincide with those in the unimodular gauge. Similarly, due to  the discontinuity at $\beta=0$ of the interpolating gauge, the correlation functions of GR in the unimodular gauge and those  in the limit $\beta \rightarrow 0$ of the interpolating gauge will not coincide. 

It is possible to derive explicit connections between VEVs of physical operators in both gauges explicitly.  Let us choose a gauge $\alpha =0$ in (\ref{S4}).  Then the action in the limit $\beta \rightarrow 0$ takes a form
\begin{multline}
S_{\text{Interpolating}(\beta \rightarrow 0)}(g_{\mu\nu}=\mu^{2/d}\hat{g}_{\mu\nu}, b^{\mu},\xi^{\mu},\overline{\xi}^{\mu}) \\
=\frac{1}{16\pi G}\int d^dx \, \omega \Big[ (R(\hat{g})-2\mu\Lambda) +b^{\mu} \hat{g}_{\mu\nu}\partial_{\rho} \hat{g}^{\rho \nu} 
-\overline{\xi}^{\mu}s \, (\hat{g}_{\mu\nu}\partial_{\rho}\hat{g}^{\rho\nu})\Big].    
\label{S57}
\end{multline}
This action coincides with that of unimodular gauge (\ref{S55}) with $\alpha=0$ and $\Lambda$ replaced by $\mu\Lambda$.\footnote{If the GR action contains terms higher orders in Riemann tensors and their derivatives, then the coefficients of those terms also depend on $\mu$.} 
 The new path integral in the interpolating gauge will be a superposition of that in the unimodular gauge with varying cosmological constant $\mu \Lambda$ and  given by 
\begin{multline}
Z_{\text{Interpolating} (\beta \rightarrow 0)}   \\
= \int_0^{\infty}d\mu\sqrt{\mu} \, \int ({\cal D}\hat{g}_{\mu\nu}{\cal D}b^{\mu}{\cal D}\xi^{\mu}{\cal D}\overline{\xi}^{\mu})_T \ e^{iS_{\text{reduced}(\beta \rightarrow 0)}(\mu^{2/d}\hat{g}_{\mu\nu}, b^{\mu},\xi^{\mu},\overline{\xi}^{\mu}, \omega)|_{\alpha=0}}
 \label{PathQGRnew} 
\end{multline}
This is a superposition of that in the unimodular gauge after Weyl transformation $\hat{g}_{\mu\nu} \rightarrow \mu^{2/d} \hat{g}_{\mu\nu}$. 
\subsection{Vacuum Expectation Values of Diff Invariant Operators in Harmonic Gauge in Terms of Those in the Unimodular Gauge}
\hspace*{5mm}
If the $\beta \rightarrow 0$ limit of the path integral of interpolating gauge is given by $\mu$ integral of that of unimodular gauge, it can be shown that VEV of physical operator $O=\int d^dx \sqrt{-g}{\cal O}(g_{\mu\nu})$ in the harmonic gauge of GR without matter fields  can be computed in terms of the $\mu$ integral of the VEV of $O$ in the unimodular gauge as follows.
\begin{eqnarray}
&&\int d^d x\langle  0|\sqrt{-g} \, {\cal O}(g_{\mu\nu})|0\rangle_{\text{harmonic}} \nonumber \\
&=&
\frac{1}{D} \int_0^{\infty} d\mu \sqrt{\mu}\, e^{-i\frac{2\mu\Lambda V_{\text{uni}}}{16\pi G}}\int d^dx \, \mu \, \omega 
\langle 0|{\cal  O}(\mu^{2/d}\hat{g}_{\mu\nu}) |0\rangle_{\text{unimodular}},
\end{eqnarray}
where
\begin{equation}
D=\int_0^{\infty} d\mu \sqrt{\mu} \,e^{-i\frac{2\mu\Lambda V_{\text{uni}}}{16\pi G}}
\end{equation}
Here $V_{\text{uni}}=\int d^d x \, \omega$ is a spacetime volume in the unimodular gauge. 
$\Lambda$ is the cosmological constant in the $\beta \rightarrow 0$ limit of the interpolating gauge.

For example in the case of VEVs of $\int d^d x \sqrt{-g}R^n(g)$, the following relation is obtained.
\begin{multline}
\langle 0|\int d^d x \sqrt{-g}R^n(g)|0\rangle_{\text{harmonic}}  \\
=\frac{(-1)^{n} 2^{2n-3}(n-2)!}{(2n-4)!}\Big(\frac{i\Lambda V_{\text{uni}}}{8\pi G}\Big)^{n-1}\langle 0| \int d^d x \, \omega \, R^n(\hat{g})|0\rangle_{\text{unimodular}} \label{Rn}
\end{multline}
Here regularization $e^{-i \epsilon \mu}$ with $\epsilon =+0$ is used in the $\mu$ integral.  If the calculation is carried out in the spacetime with Euclidean signature, `$i$' will not appear. 
Similarly VEV of the spacetime volume in the harmonic gauge is given by 
\begin{equation}
\langle 0|(\int d^d x \sqrt{-g})^n |0\rangle_{\text{harmonic}}= \frac{(2n+1)!}{4^{n}n!}\Big(\frac{8\pi G}{i\Lambda }\Big)^n
\end{equation}
Hence the spacetime volume can have an average value  in the harmonic gauge with this prescription, although the volume $V_{\text{uni}}=\int d^d x \, \omega$ is fixed in the unimodular gauge. These results explicitly show that the unimodular gauge and the harmonic gauge of GR have distinct VEVs of the same physical operators. If matters or terms which are higher-order in curvature tensors are present, there will be more $\mu$-dependent terms in the action. Problem of UV divergence and subtraction are not taken into account in the discussion of this paper. 

\section{Summary and Discussion}
\hspace*{5mm}
In this paper, it is shown that the path integral based on  the BRST invariant action for QGR in the unimodular gauge obtained in \cite{Bau1} can be simplified into a form based on functional measures based on TDiff invariant norm and coincides with that of QUG. This shows that QGR in the unimodular gauge is equivalent to QUG to all orders of perturbation theory. 

Then we studied a new BRST invariant action (\ref{S2})  in QGR which is supposed to interpolate between the harmonic gauge ($\beta \rightarrow \infty$)  and the unimodular gauge ($\beta \rightarrow 0)$. This action is obtained by adding a certain BRST-exact term to (\ref{S1}). 

It turned out, however,  that in the $\beta \rightarrow 0$ limit the action does not coincide with (\ref{S1}). In this limit the condition which should correspond to the unimodular condition is given by $\sqrt{-g}=\mu\omega$, where $\mu$ is an arbitrary positive constant, and integration over $\mu$ must be carried out. The VEVs of Diff invariant operators are the same for any gauge parameter $\beta$ except for  the value $\beta = 0$. It is shown that GR in interpolating gauge is gauge  equivalent\footnote{in the sense that VEVs of diffeomorphism operators coincide} to the superposition of GR in the unimodular gauge with varying volume element $\sqrt{-g}=\mu \omega$, (\ref{S57})-(\ref{PathQGRnew}). 
The VEVs of diffeomorphism invariant quantities in harmonic gauge will be obtained by this integration over $\mu$.\footnote{Because $\mu$ is an integration variable, the cosmological term $\mu \omega$ is not decoupled from other terms in the action.} 
This result will be also valid, even if higher-curvature and derivative terms are introduced into the action of GR. 
 
It is important to understand why a path integral for harmonic gauge of GR will be gauge equivalent to a superposition of those for unimodular gauge over those obtained by performing  constant Weyl transformations. A reason will be that in GR with harmonic gauge  the volume of spacetime $\int d^d x \sqrt{-g}$ is not fixed, while it is fixed to be $\int d^dx \, \omega$ in the unimodular gauge. So the GR theories  in both gauges will not be equal. Only by incorporating into the path integral the contributions from GR's in unimodular gauge with  volume elements of all scales, equivalence with GR in other gauges will hold. As was found in subsec.3.4 the spacetime volume can have an average value with this prescription. 

Then, because the path integral of GR in the unimodular gauge and that of UG coincide,  the following conclusion can be drawn.   A path integral for harmonic gauge of GR will be gauge equivalent to a superposition of those for UG over those obtained by performing  constant Weyl transformations on the metric tensor,  after non-dynamical cosmological term is introduced. 
This issue deserves to be studied further. 

\vspace*{2cm}

\begin{flushleft}
{\LARGE {\bf Acknowledgment}}
\end{flushleft}
The author thanks N. Ohta for useful comments on handling functional determinants which originate from  parametrization of longitudinal variations of path integral variables.



\end{document}